# Material and debris transport patterns in Moreton Bay, Australia: The influence of Lagrangian coherent structures

Kabir Suara[1*], Mohammadreza Khanarmuei[1], Anusmriti Ghosh[1], Yingying Yu[2], Hong Zhang[2], Tarmo Soomere[3], Richard J. Brown[1]

[1] Environmental Fluid Mechanics Group, Queensland University of Technology (QUT), QLD 4000, Australia

[2] School of Engineering and Built Environment. Griffith University, QLD 4215, Australia 4000, Australia

[3] Institute of Cybernetics, Tallinn University of Technology, Tallinn, Estonia

Corresponding author address: * Environmental Fluid Mechanics Group, Science and Engineering Faculty, Queensland University of Technology, 2 George St., Brisbane QLD 4000, Australia.

E-mail: k.suara@qut.edu.au

# Material and debris transport patterns in Moreton Bay, Australia: The influence of Lagrangian coherent structures

## Abstract

Coastal tidal estuaries are vital to the exchange of energy and material between inland waters and the open ocean. Debris originating from the land and ocean enter this environment and are transported by currents (river outflow and tide), wind, waves and density gradients. Understanding and predicting the source and fate of such debris has considerable environmental, economic and visual importance. We show that this issue can be addressed using the Lagrangian coherent structures (LCS) technique which is highly robust to hydrodynamic model uncertainties. Here we present a comprehensive study showing the utility of this approach to describe the fate of floating material in a coastal tidal embayment. An example is given from Moreton Bay, a semi-enclosed subtropical embayment with high morphologic, ecological and economic significance to Southeast Queensland, Australia. Transport barriers visualised by the LCS create pathways and barriers for material transport in the embayment. It was found that the wind field modified both the rate attraction and location of the transport barriers. One of the key outcomes is the demonstration of the significant role of islands in partitioning the transport of material and mixing within the embayment. The distribution of the debris sources along the shoreline are explained by the relative location of the LCS to the shoreline. Therefore, extraction of LCS can help to predict sources and fate of anthropogenic marine debris and thus, serve as a useful way for effective management of vulnerable regions and marine protected areas.

## 1.0   Introduction

Coastal waters are an ecologically rich habitat for a vast array of transient and resident species. The waterways play an important role in the exchange of energy and material between inland waters and the open ocean [*Geyer & MacCready*, 2014]. In economic terms, some of these waters provide sheltered deep-water port access, shorelines for fish breeding and support for the functioning of the entire marine ecosystem, ground water discharge, effluent disposal, fisheries/aquaculture, tourism and recreation. Any of these functions can be severely impacted by catchment decisions, e.g. the control of river discharge, that affect the transport of water and material [*Wolanski*, 2015]. In addition, mitigation and management of pollutant and debris rely on realistic assessments and predictability of the dispersal of these materials [*Fischer et al.*, 1979]. Therefore, continuous improvement of the predictability of the fate of material in coastal waters is important for keeping sustainability of the relevant ecosystem [*D'Asaro et al.*, 2018]

The horizontal spreading and mixing processes of various substances such as floating debris, pollutant spills, and plankton patchiness in estuaries are impacted by underlying flow dynamics. In time-dependent, chaotic flow, an Eulerian viewpoint provides limited information about the transport of material, therefore, it is natural to approach the problems of material transport in a Lagrangian framework [*Vandenbulcke et al.*, 2009; *Wei et al.*, 2018]. Lagrangian particle tracking is applied to examine the fate, source and mixing of materials in many geophysical flows [*Fredj et al.*, 2016] including oil and debris in the near shore [*Suneel et al.*, 2016; *Qiao et al.*, 2019]. In addition, significant emphasis has been aimed at managing the floating marine debris through intensive shoreline clean-up in different parts of the world [*Hardesty & Wilcox*, 2011; *AMDI*, 2018; *Yu et al.*, 2018; *Chen et al.*, 2019]. While these shore-based studies may provide some approximation of the composition and abundance of debris in adjacent waterways, they represent only the fraction of debris in the adjacent waterways [*Hardesty & Wilcox*, 2011]. Therefore, a more comprehensive understanding and management of debris would include incorporating modelling approaches that can chart the likely path of material as well as at-sea surveys, e.g. [*Carlson et al.*, 2017].

Theoretical work on dynamical systems has characterised transport barriers in unsteady flows as Lagrangian coherent structures (LCSs) [*Haller*, 2001]. These structures are locally attracting/repelling material lines, that accumulate/spread materials, acting as the core of the Lagrangian transport pattern and form the major transport pathways [*Olascoaga et al.*, 2006]. The LCS concept has been used to understand a range of processes and problems in oceanic flow [*d'Ovidio et al.*, 2004; *Prants*, 2014]. Coastal tidal flows have recently been a focus of studies using LCSs [*Huhn et al.*, 2012; *Wei et al.*, 2018; *Lacorata et al.*, 2019]. These flows are particularly challenging due to complex three-dimensional boundaries, the interplay of unsteady forcings of wind, tide and river discharge as well as their interaction with physical features, including bathymetry and island structures [*Suara et al.*, 2017]. How these different factors modify the pattern of transport pathways in a tidal coastal embayment is yet not well understood.

In unsteady coastal waters, 'perfect' barriers to transport are uncommon and exist within a relatively short timescale e.g., tidal inlet jets and fronts. The transport barriers identified in this context are relative to other parts of the flow field. This diagnostic approach, ridges of high finite-time Lyapunov exponent (FTLE) values, which visualises potential LCSs and identifies parts of the flow field that accumulate/spread materials more than others is applied. More robust and exact methods for identification of LCS are in the literature [*Haller*, 2011]. A similar FTLE method was applied to identify transport barriers in coastal waters using HF radar derived surface velocity [*Lekien et al.*, 2005; *Coulliette et al.*, 2007], and numerical model velocity [*Huhn et al.*, 2012; *Wei et al.*, 2018]. The concept is applied here to identify transport barriers in a tidal embayment using the Eulerian velocity obtained from a hydrodynamic model velocity field and to examine the role of wind and tides on transport barriers in such a system [*Yu et al.*, 2016].

Moreton Bay (MB) is a semi-enclosed subtropical embayment high in morphologic, ecological and economic significance to southeast Queensland, Australia [*Gibbes et al.*, 2014]. The bay lies between 27º and 28º south latitude, spans approximately 110 km north to south, and has its major opening to the ocean of approximately 15 km on the northern end (Figure 1). The system holds the discharge from many smaller rivers and estuaries [*Suara et al.*, 2015], as well as the large Brisbane River, and is thus prone to strong changes in physicochemical qualities due to anthropogenic activities and stormwater discharges. It also serves as a natural protection for a section of Australia's east coast from direct wave action. In recent years an increasing number of severe flood events have resulted in large sediment transport in the bay [*Brown & Chanson*, 2013; *Yu et al.*, 2014a]. The sub-tropical climate of MB is characterised by high rainfall during summer months that can lead to large runoff events and occasional floods, while the base-flow is minimal during the winter dry season [*Gibbes et al.*, 2014]. For most of the time, MB receives wind from the southwest and the northeast directions in the morning and afternoon, respectively [*ABM*, 2018].

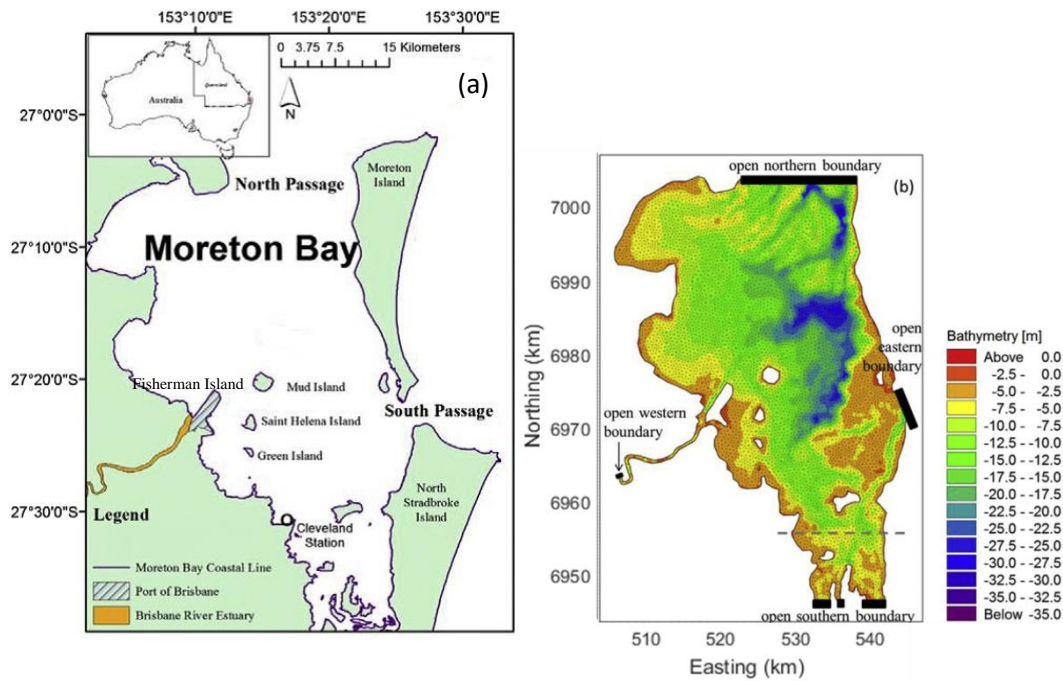

Figure 1: (a) Map of Moreton Bay showing the main features (shoreline, islands and ocean passages) (b) The bathymetry of Moreton Bay [*Yu et al.*, 2016].

The aim of this work is to show how the identification of material lines can improve our understanding of the circulation pattern and fate of debris in a tidal embayment. The strongest attracting and repelling material lines dictate the global mixing pattern and effectively separate masses of water of an unsteady flow at a finite time interval. These material lines which can be visualised by LCS therefore act as transport barriers. The dynamics of the transport barriers with tidal phase and type are investigated. Furthermore, the effect of wind on LCS is examined through several levels of submergence of floating particles in the tidal flow. Section 2 presents the diagnostic tool utilised for the identification of the LCS from velocity output of a validated hydrodynamic model of MB [*Yu et al.*, 2014a; *Yu et al.*, 2016]. Section 3 presents the results and discussion and Section 4 summarises the key outcomes.

## 2.0  Methods and data processing

### 2.1 The hydrodynamic model

The LCS analyses are carried out on the surface velocity field produced from a hydrodynamic model of Moreton Bay (MB). Please refer to *Yu et al.* [2016] for a detailed model description including model validation. The MB hydrodynamic model was developed using MIKE3D, which has been widely used in studying dynamics of estuaries and coastal waters [*Yu et al.*, 2014b; *Yu et al.*, 2016]. The model solves the three-dimensional Reynolds averaged Navier–Stokes equations with the assumptions of Boussinesq and of hydrostatic pressure [*Yu et al.*, 2016]. The computational domain (Figure 1) was modelled as a network of unstructured triangular grids consisting of 13,918 elements in the horizontal plane. Grid size near the Brisbane River mouth was <100 m while the size ranged 100–500 m in the other parts of the domain. The depth of the model varied from 0 – 35 m, where 0 corresponds to the mean sea level (Figure 1b) The vertical direction was modelled using a variable thickness sigma-coordinate system consisting of 10 layers of thickness ~0.5 m near the boundaries (surface and bed) and 2–3 m in the middle depending on the depth. The western boundary (Figure 1b) was imposed with hourly river discharge data derived from observations by the Department of Environment and Resource Management, Queensland [*Yu et al.*, 2016]. The model was forced by prescribed tidal elevations in the northern, eastern and southern boundaries which are connected to the ocean. Wind data from Australian Bureau of Meteorology [*ABM*, 2018] were used as model wind input. Monthly average temperature and salinity were used as the initial conditions while the model was spun-up for a month period prior to the intended model period [*Yu et al.*, 2014a; *Yu et al.*, 2016]. This is to ensure that a dynamic steady state was reached, and the output is not affected by the initial conditions.

The hydrodynamic model was calibrated and verified using field observations [*Yu et al.*, 2014b; *Yu et al.*, 2016]. The model generally produced accurate results of the flow fields. The root mean square error (RMSE) of the water level was a maximum of 7 cm. The normalised RMSE between the observed (drifter) and modelled trajectories were 1.26% and 7.45% in the northing and easting directions, respectively [*Yu et al.*, 2016]. The RMSE of the simulated water temperature and salinity were 0.9 °C and 0.23 psu and are presented in details in [*Yu et al.*, 2016].

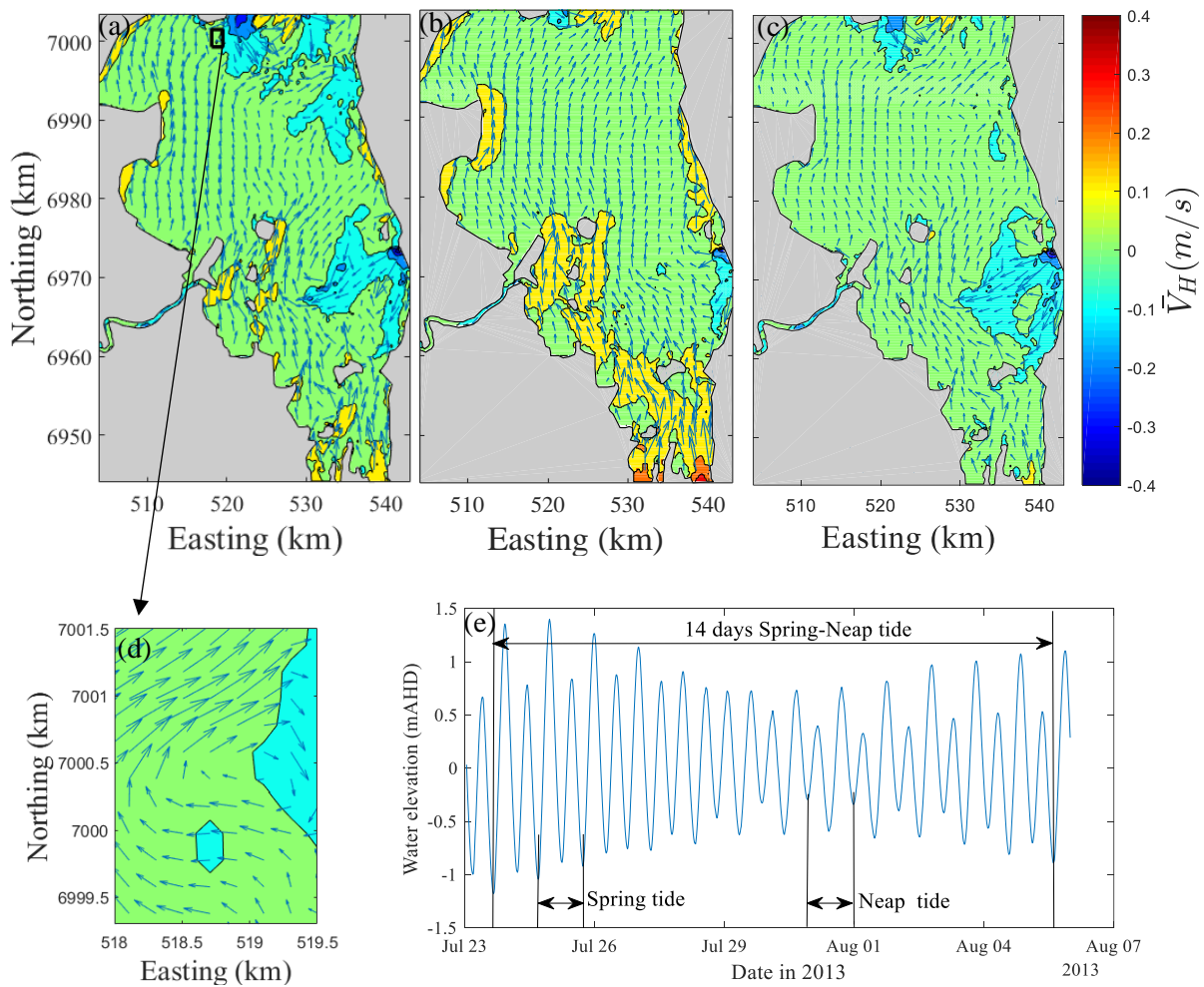

Figure 2: Tidal circulation pattern of Moreton Bay shown as average horizontal velocity vector coloured by its magnitude and the direction (positive towards the north) for a typical, (a) Spring tide, (b) Neap tide, (c) 14 days spring-neap average. Note that each vector represents a cluster of 8 grid point for clarity. (d) The blown-up inset black box in the northern end in (a) showing secondary flow (e) Time series of the water elevation showing the time range for the averages in (a) – (c).

The particle tracking and the LCS analysis here were carried out using the surface velocity output (top layer) with a 15-minute time interval between 23 July and 5 August 2013, interpolated into a structured cartesian two-dimensional grid of 200 m spacing. The period was chosen because it coincided with a period of intensive fieldworks which was used in for validation purposes [*Yu et al.*, 2016]. The 14 days output covered spring and neap tidal types. Figure 2 shows the averaged flow during typical spring tide, neap tide and 14-day spring-neap average. Because of the flood and ebb pattern of the tidal flow, Figure 2 uses the horizontal velocity magnitude and the Northing direction (positive towards the north). The figure shows that the circulation pattern in MB was such that the major transport is northward. Also, localised voritcal structures behind the islands were prominent structures within the middle and the southern parts of the embayment.

### 2.2 Lagrangian Coherent Structures from hydrodynamic model

In this analysis, the finite time Lyapunov exponent (FTLE) is employed as a diagnostic quantity for the LCSs. This diagnostic approach among others is known for the simplicity and objectivity of its algorithms and serving as proxies for LCSs [*Peikert et al.*, 2014; *Hadjighasem et al.*, 2017]. Comparison across different detection methods is presented in detail in *Hadjighasem et al.* [2017]. Here, the Lagrangian particles are passively advected by the flow field and trajectories and are obtained by solving the Lagrangian fluid particle motion using:

$$\dot{x} = u(x,t),  \quad\quad\quad 1$$

where $u(x,t)$ is the prescribed flow field, that is, the surface velocity output of the hydrodynamic model described in Section 2.1. The calculation is restricted to the two-dimensional surface velocity which is valid for buoyant particles that are locked to the water surface. The flow map $\Phi_t^{t+T}$ takes the fluid particles from the locations at time $t$ to their final locations at time $t+T$ such that:

$$\Phi_t^{t+T} : x(t) \longrightarrow x(t+\tau)  \quad\quad\quad 2$$

The FTLE field can be obtained from the flow map such that it represents the finite-time average of the maximum expansion or contraction rate of the pairs of particles, $\lambda_{max}$ advected by the flow map such that:

$$FTLE(x_0, t_0) = \frac{1}{|\tau|} \ln \sqrt{\lambda_{max}(\Delta)},\qquad\qquad 3$$

where $\lambda_{max}$ is the largest eigenvalue of the Cauchy–Green deformation tensor $\Delta = \nabla^*_{x_o} \nabla_{x_o}$, $\nabla_{x_o} = \partial \Phi(t_1, x_o, t_o)/\partial x_o$ is obtained from the first order spatial derivative of the flow map at the final trajectory positions with respect to the initial positions and $\tau = t_1 - t_0$ is the integration time while indices 0 and 1 indicates the initial and the final times. The line of the FTLE field which intuitively is the curve that goes through the local maxima of the FTLE evaluated along the direction of fastest change in the FTLE signals a potential LCSs [*Haller*, 2001]. This is readily identified from visual examination and thresholding of the FTLE contour plots [*Rockwood et al.*, 2019]. However, a more precise definition and extraction procedures for LCS are provided in *Haller* [2011]. Particles are advected forward ($\tau > 0$) and backward ($\tau < 0$), separately, in time to reveal the repelling/stable and the attracting/unstable material lines, respectively. Repelling (stable) material lines represent lines of maximum spreading, so initially proximal particles separate rapidly. The attracting/unstable material lines represent those with maximum accumulation. Forward time integration of particle trajectories and the ensuing FTLE field are useful for addressing problems that focus only on particle fates. Combined information from backward and forward time particle integrations is useful to identify both transport pathways and fates of particles [*Lipphardt et al.*, 2006]. In the same vein, combined forward and backward FTLE fields, are useful to visualise transport pathways and fate of materials.

The calculations of the FTLE field were completed using a set of MATLAB algorithms developed in-house following the procedure below [*Shadden et al.*, 2005]. Neutrally buoyant tracer particles are seeded into the fluid domain at $t_0$. At each time step, the algorithms solve Equation (1) using the Runge–Kutta (4, 5) method, i.e., the MATLAB ode45 solver with an absolute integration tolerance of $10^{-6}$, for the particles to get their trajectories and final locations at $t_1$. A linear interpolation scheme is employed in space and time to obtain the velocities along particle trajectories. A free slip boundary condition is employed and the calculation of the trajectory of particles is stopped once they exit the flow domain through the land or ocean boundaries. Because the available velocity field does not include areas outside MB, particles that exit the domain cannot be tracked, resulting in a shorter integration time, $\tau$ for those particles. Therefore, to ensure consistency of the calculation of FTLE, Eq (3) is evaluated at the point of exit of the particles that cross the boundaries using $t_1$ as the time of exit [*Lekien et al.*, 2005]. These FTLE equation is evaluated at the maximum integration time for particles remaining in the domain. The deformation tensor is computed using finite central differencing of the separation with the neighbouring tracer particles [*Coulliette et al.*, 2007].

In the calculation of the FTLE field, the initial tracer particle density and the maximum integration times are of key consideration [*Huhn et al.*, 2012]. The statistics of the FTLE field is examined as a function of the FTLE resolution, i.e. the initial tracer particle density. A refinement factor (RF) which is the ratio of the velocity field grid to that of the FTLE is introduced. Figure 3a shows the PDF of the FTLE field for different refinement factors (RF). There was no significant difference in the distribution of the FTLE field for RF ≥ 4. The spatial average of the FTLE field becomes independent of the ratio of the FTLE and velocity field resolution after a factor of two whereas the maximum FTLE grows with the refinement of the FTLE field (Figure 3b). Therefore, an initial spacing of 50 m is employed, i.e., a resolution four times higher than that of the velocity field. Further increase in the spatial resolution of the tracer particles increased the FTLE maximum, thus refining the LCS, but without revealing new LCS information from inspection of the level sets. An integration time, $\tau = 72$ hours is selected such that it is shorter than the average residence time of MB (>30 days) and longer than the smallest process of interest, i.e., M2 tidal period [*Yu et al.*, 2016; *Ghosh et al.*, 2018]. Particles that left the domain before the 72 hours constitute about 25% of the initial seedings.

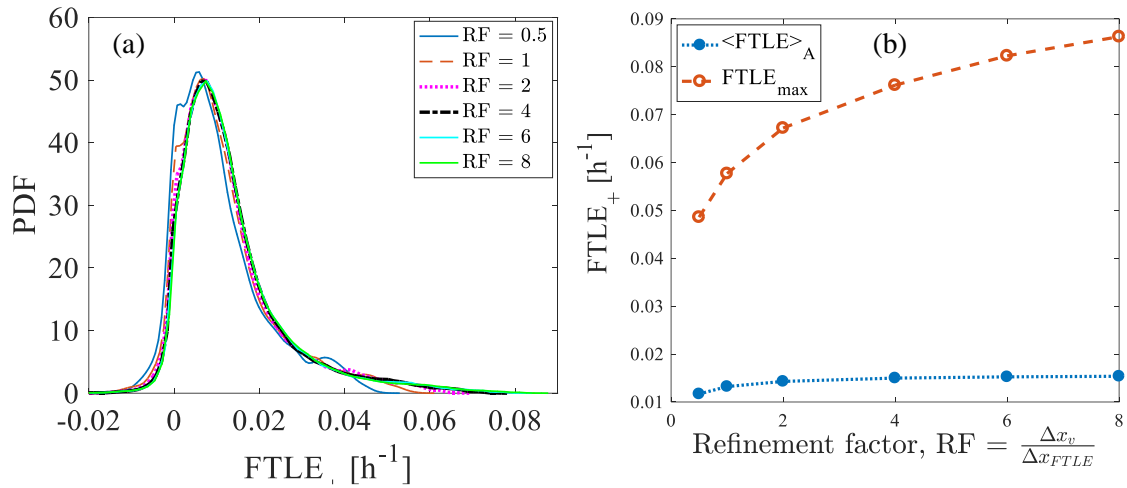

Figure 3: Statistics of the FTLE field (a) Probability distribution function for different FTLE field resolution, (b) Spatial average FTLE field and maximum value of the FTLE field as a function of refinement factor. $\Delta x_v$ is the velocity grid size and $\Delta x_{FTLE}$ is the FTLE grid size.

### 2.3 Windage effect on particle statistics

Floating materials do not necessary move with the water surface due to additional forces from wind and waves, therefore this modified transport needs to be accounted for in relevant LCS analysis [*Allshouse et al.*, 2017]. Trajectories of floating tracer particles are directly affected by the combination of current, wind and wave properties. These factors potentially influence the estimate of the FTLE field and the ensuing hyperbolic LCS structures. MB is a semi-enclosed system, sheltered from direct impact of ocean waves and associated Stokes drift, therefore, the impact of the waves is assumed to be negligible. *Allshouse et al.* [2017] examined the impact of wind on LCSs in open coastal waters and showed that the impact of wind cannot be known a *priori* and recommended that windage should be accounted for in realistic applications of LCSs. For this reason, the impact of windage on the Lagrangian trajectories and the FTLE estimates in MB is further examined. To account for the windage effect, a hybrid velocity field is introduced into Eq. (1) such that:

$$u(x,t) = u_c(x,t) + C_w u_w(x,t),  \qquad 4$$

where $u_c(x,t)$ is the water surface velocity field from the hydrodynamic model, $u_w(x,t)$ is the wind field adjusted to the standard 10 m height above the sea level and $C_w$ is the windage coefficient. Although, there are several aspects of windage that can affect the hybrid velocity field, the orientation and the level of submergence of floating material are the two dominant considerations [*Yoon et al.*, 2010; *Hardesty & Wilcox*, 2011]. A simple linear model for the windage is employed such that:

$$C_w = k\sqrt{\frac{A}{W}}, \qquad 5$$

where $A$ is the total cross-sectional area exposed in air, $W$ is the total cross-sectional area submerged in the water, $k = 0.03$ is the wind pressure from an empirical relation following [*Allen*, 2005; *Yoon et al.*, 2010]. Plastic materials contribute to about 72% of the debris collected around Australian shores in 2018 [*AMDI*, 2018]. Here we consider $A/W = 0$, 0.1 and 0.5 which is representative of most floating debris. $A/W = 0$ represents debris that are neutrally buoyant, floating just below water surface. $A/W = 0.1$ represents debris with 9.1% of their surface area above water, such as plastic pieces [*AMDI*, 2018] while $A/W = 0.5$ represents those with 33.3% of their surface area above water such as plastic floating lighters [*Yoon et al.*, 2010].

The wind field was created by spatial interpolation of wind measurements from 13 weather stations distributed in and around MB and described in detail in the Supplementary material. Figure 4 shows the sample wind, flow and hybrid fields (Eq. 4) for $A/W = 0.1$ and 0.5. The locations of the weather stations relative to MB and an animation over a 48-hour window for each windage are provided in the Supplementary Material (SV1 and SV2). The wind field is mainly dictated by the observations from weather stations within close proximity of the MB. The far field stations were only included to avoid errors that could result from extrapolation. It can be observed that there is a strong spatiotemporal variation in the wind field in MB indicating a level of importance in the transport. In addition, the level of submergence dictates the deviation of the hybrid field from the surface flow field.

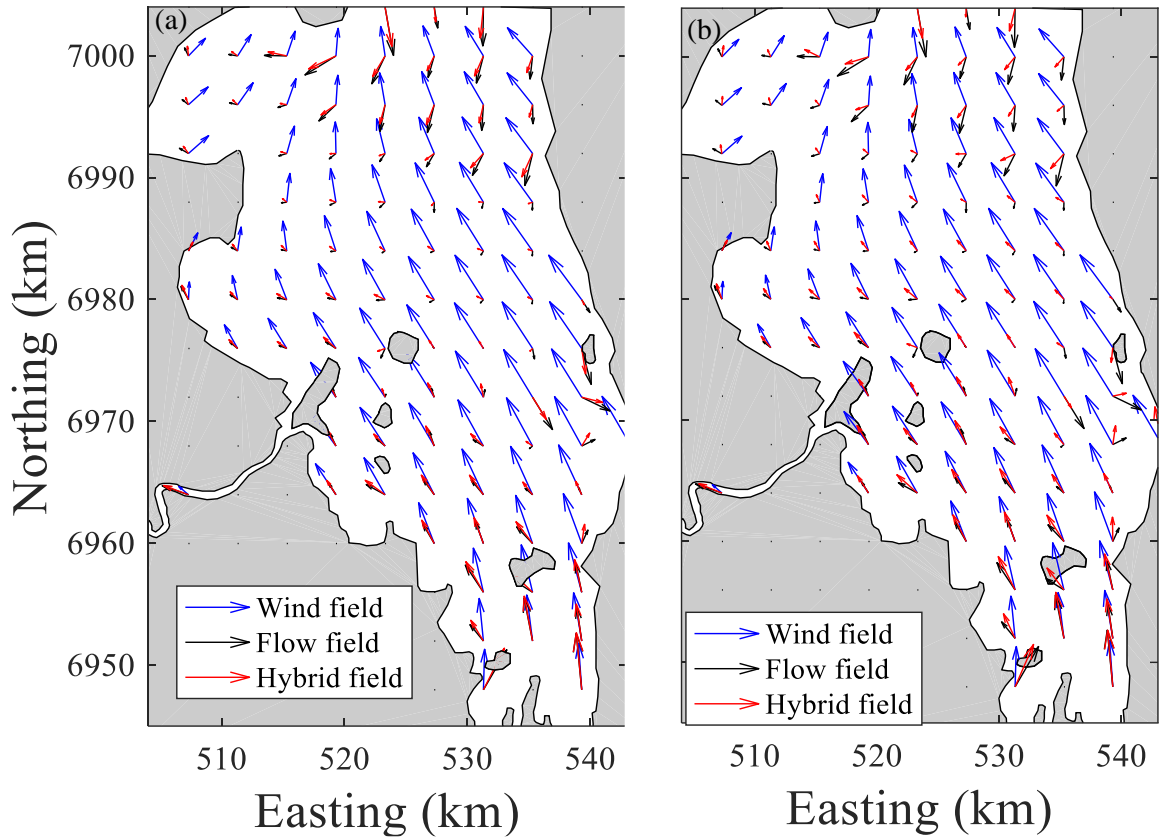

Figure 4: Typical wind field at 10 m above the mean sea level (blue arrows), water surface flow field (black arrows) and hybrid velocity field obtained from Eq. 4 (red arrows): Animations of these figures over a 48-hour window are provided in the Supplementary Material. (a) $A/W = 0.1$ (b) $A/W = 0.5$; July 29, 2013 at 23:48; Ebb tide.

## 3.0 Results

### 3.1 Transport barriers in Moreton Bay and island wakes

As described in Section 2, the FTLE fields are obtained from advecting tracer particles with the Eulerian velocity field obtained from a hydrodynamic model of MB. Figure 5 shows examples of the FTLE contour with a forward and backward integration time $\tau = 72$ h, superimposed with the corresponding residual circulation. The residual circulation represented by the residual flow field is obtained by time averaging the velocity field over the duration of the FTLE field. The residual circulations show that the flow in the vicinity of the Northern

Passage was significantly larger (> 0.5 m/s) compared to the Southern Passage and other parts of MB. This indicates that the major oceanic exchange occurs in the Northern Passage. The instantaneous and residual flow fields also highlight the presence of distinct vortical structures (~2 km length) behind the islands (e.g. Mud and Peel Islands of MB (Supplementary video SV5 and residual vectors in Figure 5). Some small but distinct vortical structures (~0.5 km) were also noticeable between islands in the south eastern boundary. The presence of these structures at long advection time $\tau = 72$ h suggests that they are unsteady with reasonable turn over time and are sustained by the northward net transport within MB. This also suggests that the transport of material in MB is strongly connected with the interactions between the tide and the wakes of the basin scale structures such as the islands and the shoreline irregularity marked by headlands.

In Figure 5(a) and (b), the FTLE fields highlight the prominent repelling (stable) material lines in red and the attracting (unstable) material lines in blue for $\tau = 72$ h, respectively. These lines indicate potential pathways for material transport by spreading (Figure 5a) and accumulating (Figure 5b). Maximal and minimal FTLE values were around 0.08 h$^{-1}$ corresponding to the exponential expansion and contraction, respectively, at a time scale equivalent to a semi-diurnal tidal cycle, 12 hours 25 minutes. Fully developed eddies were absent within the bay but filament structures which are coherent up to the integration time appear in the wake of the islands and sharp headlands. These type of motions which are strongly circularly polarised and reveal numerous instantaneous eddy-like structures but do not contain long-living eddies often occur in estuaries of comparable size and with complex geometry and bathymetry (*e.g.[Veneziani et al., 2005; Viikmäe et al., 2013]).* Unlike the work of *Huhn et al.* [2012] where LCS were not pronounced in the middle of Ria de Vigo estuary, the LCS in MB are attached to shorelines and islands with the majority attached to islands. The LCSs fold and stretch from these points of attachment into the flow depending on the integration time. An important result from observing the dynamics of LCS in MB is that while stable material lines are attached to both the shoreline and the islands, the unstable material lines are radiated predominantly from the wakes of the islands. This suggests that the presence of the material lines depends on the relative location of the islands, the magnitude of the flow and eddy decay time scales within the bay.

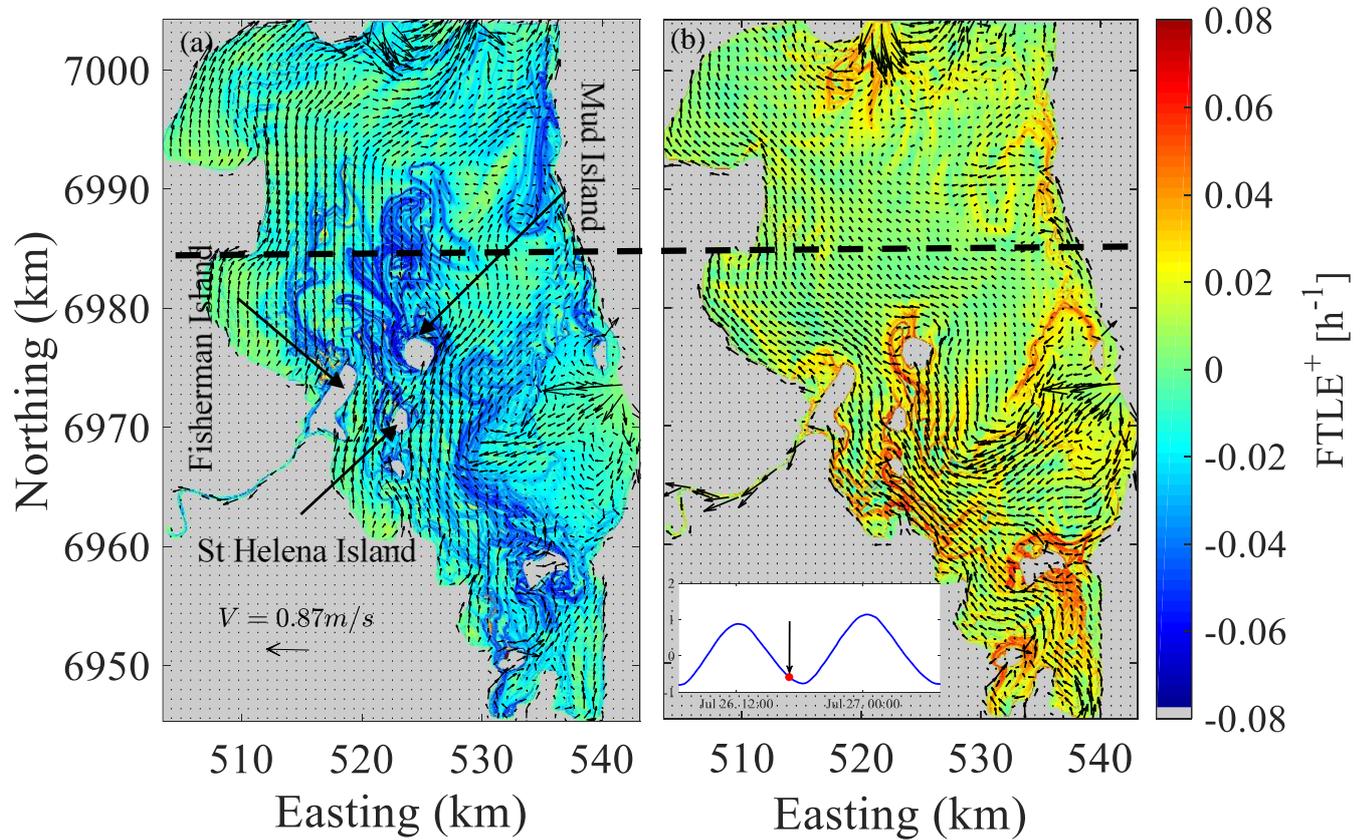

Figure 5: Example of the FTLE fields overlaid by the residual flow field ($\tau = 72$ h): The thick dash line represents the boundary of the northern and southern part discussed in text. (a) Backward FTLE field, (b) Forward FTLE field. The colorbar is relevant to both (a) and (b). Note that the backward FTLE field is multiplied by negative 1 for visualisation.

*Wolanski et al.* [1984] describe the balance of eddy generation with the destruction by vertical diffusion behind an island through an island wake parameter, P, such that:

$$P = \frac{UH^2}{K_z L}, \qquad 6$$

where $U$ is the velocity of the flow, $K_z$ is the vertical eddy diffusivity, $L$ is the characteristic length of the island and $H$ is the average water depth. This parameter describes the formation and dynamics of island wakes. Values of $P < 1$ signify an absence of wake structure, while values of $P$ close to 1 indicate the presence of a steady wake. When $P$ is above the limiting value of 1, the island wake is unsteady, eddies are advected away from the island and the wake behaves as that formed at high Reynolds number [*Wolanski et al.*, 1984]. The islands within MB have characteristic as follows; $L$ = 1.2–4.02 km, $H$ ~ 10 m, $U$ ~ 0.5 m/s and assuming $K_z$ = 0.001 m$^2$/s – the upper vertical eddy diffusivity limit in the ocean and a value larger than those observed in a tidal estuary of similar depth [*Gargett*, 1984; *Etemad-Shahidi & Imberger*, 2005]. This results in an estimate of $P$ ranging from 12 to 50. This further indicates that the presence of material lines attached to the islands in MB result from the unsteady wakes generated by these islands.

### 3.2 The Tidal variation

The dynamics of the MB flow is represented by the time dependent FTLE field shown as snapshot images of one-hour interval over a 12-hour period for a typical spring tide (Figure 6). In order to meaningfully compare the FTLE field at different times, the results are normalised by the maximum FTLE values for the forward and backward fields, separately. A hyperbolic FTLE field = FTLE$^+$ – FTLE$^-$ is defined, combining the forward and backward fields following *d'Ovidio et al.* [2004]. This visualisation approach suffers some deficiency in that there is tendency for reduction in the resulting values in areas where stable and unstable material lines coincide. However, it provides a simplified approach for visualising potential hyperbolic LCS from the FTLE field and identifying the dominant hyperbolic saddle points. The saddle points are locations where lines of stable (red) and unstable (blue) material lines meet. They are the main hidden features that dictate mixing and stirring of tracers in the flow field.

The normalised FTLE fields (Figure 6) showed that dynamics of the northern and southern parts of MB are different. In the northern part, the material lines are more distinctly separated and reveal different pathways to material transport as a function of their initial location relative to the islands. It is therefore likely that masses of adjacent water bodies with higher likelihoods of different physicochemical properties often occur. On the other hand, the southern part has material lines that are closely packed. This feature apparently results from the proximity of the islands to the shoreline and to one another. The associated dynamics of water masses involves strong stirring of massless neutrally buoyant tracers. However, the presence of vortices can cause selective dispersion of non-neutrally buoyant inertial particles [*Chanson & Tan*, 2011; *Pinton & Sawford*, 2012].

Because of the arrangement of the stable and unstable material lines, the northern part of MB has fewer numbers of saddle points than the southern part of the bay. The main saddle points in the northern part are located next to the northern ocean boundary resulting from material lines attached to the northeast of the study area. Despite the presence of large oceanic exchange through this boundary, the presence of saddle points in this location is likely related to factors including vertical stratification due to its proximity to the inlet and strong variations of the water depth (Figure 1b) that can cause divergence in such a quasi-two dimensional flow field [*Suara et al.*, 2017]. Contrariwise, the dynamics of the southern part is governed by more pronounced saddle points. These multiple saddle points are created by material lines radiating from both the shoreline and islands and crossing in close proximity. The formation of the complex structure can be explained by the ratio of tidal excursion length and the length scale for material transport between the physical barriers (islands and headlands) in the flow. The residual tidal current in MB varied between 0.17 and 0.61 m/s, [*Pattiaratchi & Harris*, 2003; *Yu et al.*, 2016] corresponding to tracer displacement varying between 3.2 – 13.2 km within a single flood or ebb tide [*Pattiaratchi & Harris*, 2003; *Yu et al.*, 2016]. However, the separation between the islands in the southern part is small (<3 km, e.g., Mud Island and St. Helena Island,

Figure 6). Therefore, eddies generated and propagated during a phase of the tide as a result of the unsteadiness in the wake of the island are more likely to encounter another island or shoreline edge before the turning of the tide.

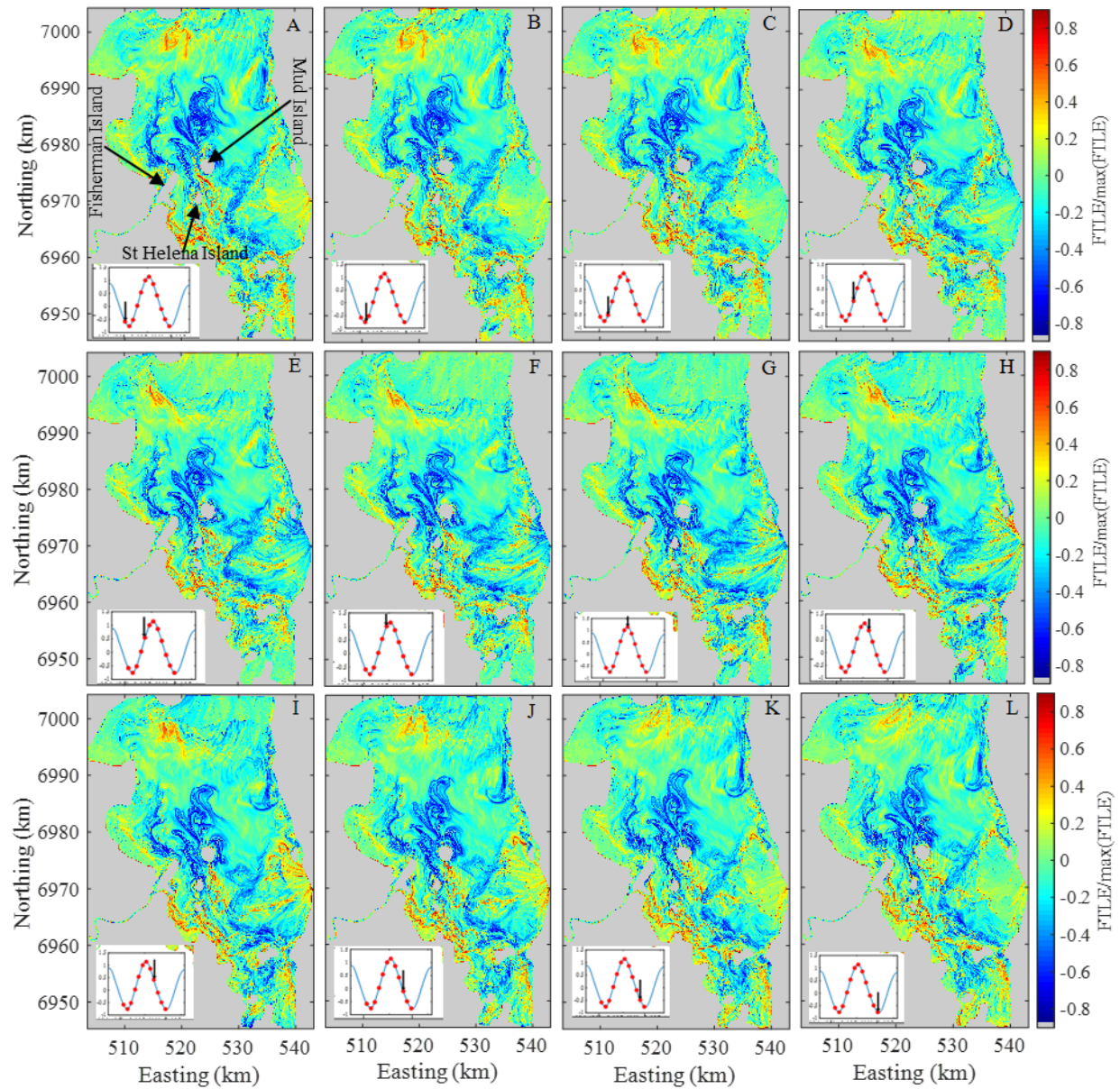

Figure 6: Normalised FTLE field showing the repelling and attracting structures during a spring tide as a function of tidal phase: A–L indicate the initial tidal phase from around low tide to the next low tide with a step of approximately $\pi/6$. The water level at the initial time, $t_0$, are shown as inserts on the left lower corner.

The flow structure and the location of these saddle points varied as a function of tidal phase. For example, at low tides (e.g. Figure 6G), the stable material line (red) apparently connected the northeast shoreline to the east. This structure folds and stretches with the tidal phase creating multiple saddle points as it intersects the unstable material lines (blue), radiating from the wakes of the islands. The stable material lines stretch during the rising tide and reverse during the falling tide. The opposite is observed with the unstable material lines. Similarly, the point of attachment of the stable material lines (red) on Mud Island rotates counter clockwise approximately 180º, starting from the north in A, within the 12 hour snapshot (Figure 6).

### 3.3 Effect of the wind on LCS

Transport of floating material in the surface layer of water bodies is known to be significantly modified by wind [*Yoon et al.*, 2010; *Hardesty & Wilcox*, 2011]. One of the strongest attributes of LCS analysis is its robustness to noise or uncertainties in the velocity field [*Haller*, 2002; *Allshouse et al.*, 2017]. However, wind distribution, in a semi-enclosed embayment of complicated shape, particularly for a short time-scale is far from being random and therefore, the impact of the wind field is expected to modify transport in the system. *Allshouse et al.* [2017] investigated the impact of windage on a hyperbolic saddle point influencing the transport of material in an open coastal water coral reef at Ningaloo Western Australia. They concluded that the impact of wind is important and cannot be known *a priori*. Here the impact of windage on the transport barriers in a tidal embayment MB is examined using a realistic wind field. For this calculation, windage is accounted for in the Eulerian flow field using Eqs. 4 & 5 before being used in Eq. 1. The wind predominately blew from the southwest for the duration covered by the analysis, which is typical of the winter season in MB.

LCS are shown as the ridges of the FTLE field in Figure 7. The ridges are rendered by showing larger 50% of the maximum FTLE values. We compare LCSs calculated from tracer particles seeded with two different areas above water with those driven by only the water surface flow field as shown in Figure 7(a, b & c). The inclusion of the windage modified both the locations and the rate of attraction/repulsion of the barriers highlighted by the ridges of the FTLE field.

One of the discernible impacts of the windage is that it stretches out the material lines that were connected to the islands in MB. For example, in the northern part of the bay, the unstable material lines attached to Fisherman Island, Mud Island and St Helena Island were stretched out northward with the windage. This stretching connected the unstable material lines to the northern ocean boundary and the northern shoreline for $A/W = 0.1$ and $0.5$, respectively. The windage increased the maximum FTLE field, which implies an increase in the rate of expansion and contraction along the trajectory of the material lines for materials adjacent the stable and unstable material lines, respectively. On the other hand, the stretching of the material line reduced the density of the hyperbolic saddle points. The number of points where the normalised forward and backward FTLE values were simultaneously obtained reduced by 30% and 50% for $A/W = 0.1$ and $A/W = 0.5$, respectively. This can be seen for example, in the southern part where hyperbolic saddle points are spaced out with the increase in the windage and potentially reducing mixing/stirring processes in the embayment.

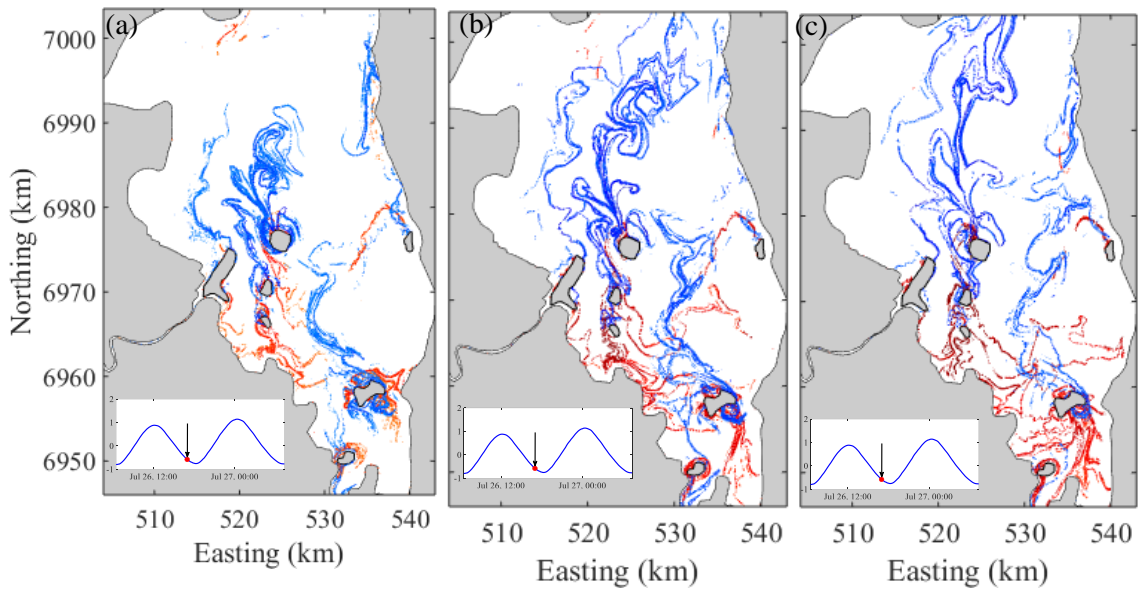

Figure 7: LCS shown as ridges of the FTLE field; red lines renders the stable material lines (spreading lines) while blue line renders the unstable material lines (accumulating lines) (a) for $A/W = 0$; (b) $A/W = 0.1$ (c) $A/W = 0.5$; Note for visualisation, the white spaces have FTLE values lower than or equal to 50% of the FTLE maximum

In summary, the windage effect on the LCS in MB for partly submerged materials is such that it (i) increased the rate of their contraction and expansion in the vicinity of the barriers,

(ii) reduced the mixing/stirring processes, and (iii) modified the direction of their transport as unveiled by the barriers. The modification in the direction of transport is consistent with the results in the Ningaloo Reef, where wind modified the trajectory of dominant hyperbolic saddle points [*Allshouse et al.*, 2017]. The results here are representative of real wind conditions in the MB tidal embayment for wind predominantly blowing from the southwest (i.e., towards the northeast). The wind from other directions which are not covered here, are likely to impose additional modification to the properties of the LCS in MB.

The current results however showed that wind effects on the LCSs have a potential to influence the fate of pollutants and other passive objects. For example in the case of $A/W = 0.5$, the presence of the extended unstable material lines that are connected to the northern shoreline would result in an increase in the residence time in MB because of reduction in the exchange of material with the ocean and increase in the shore hits for floating materials coming from the western shoreline. This effect is discussed in further detail in the next Section in which visualised LCSs are linked with the sources and the fates of floating materials in the shore. The analyses here indicate that the visualised transport barriers are affected by windage and the application of the concept to pollution management requires a proper consideration of the wind among other effects such as inertial size, which is not covered in this work.

### 3.4 Sources and fate of debris in the shoreline

#### 3.4.1 LCS and source distribution for debris clean-ups

There is a significant effort on coastal water clean-up in Australia particularly along the Queensland shorelines [*AMDI*, 2018]. It is therefore important to examine the role of LCSs on the source and the fate of debris in an embayment such as MB. This task has a practical implication to assist in identifying areas with relatively higher chances of debris wash-up for managing the limited resources available for such activity. The Australian Marine Debris Initiative is a non-government organisation that holds the database of most clean-ups along Australian shoreline [*AMDI*, 2018]. AMDI provides standard unified clean-up procedures for volunteers and organisations for categorising debris collected and for reporting into the database. This categorisation allows sorting debris into different sources to assist in a Source Reduction Plan (SRP) development that aims at a reduction of debris directly from the source. The categorisation also allows differentiating between debris from the land (i.e., from inland waters) from those originating from the sea. For example, plastic containers and food packaging materials are categorised as originating from the land while fishing debris are categorised as originating from the sea.

To examine the role of the LCS in determining the sources of debris, the contributions of the land and sea-sourced debris to the total number of debris collected along four different sections of the western shoreline of MB between 2011 and 2018 are examined (Figure 8). These sections are selected because they are next to major cities with the highest population along the MB shoreline and their proximity to the Brisbane River which holds the largest discharge into MB and major shipping fairways to the Brisbane Port within MB. Such locations have the tendency to release higher pollutants, debris and stormwater-driven substances into the bay.

Figure 8 shows the distribution of marine debris collected along the shoreline in Deception Bay (DB), Redcliffe Headland (RCH), Bramble Bay (BB) and Waterloo Bay (WB) over a period of 8 years (2011-2018). The dataset is extracted from Marine Debris Database [*AMDI*, 2018]. The areas of the pie charts correspond to the scaled total number of debris collected at the selected shoreline sections marked. The contributions of the sea-sourced debris for the DB, RCH, BB, and WB were 32%, 16%, 7% and 26%, respectively. The figure suggests that the relative location of the LCSs to the shoreline affects the source distribution with the LCSs acting as a transport barrier that shapes the contributions of debris from land and ocean.

For example, DB which had the least obstruction from the LCSs and largest clearance to the Northern Passage, had a significant portion (32%) of debris originating from the sea. On the other hand, RCH and BB were obstructed because of their location between the shoreline and the unstable material lines radiating from the Fisherman Island. It is likely that the presence of transport barriers favours retentions, i.e. reduces the exchange of land-sourced debris with the oceans. This may partly explain why more than 80% of the debris collected in BB and RCH are land-sourced.

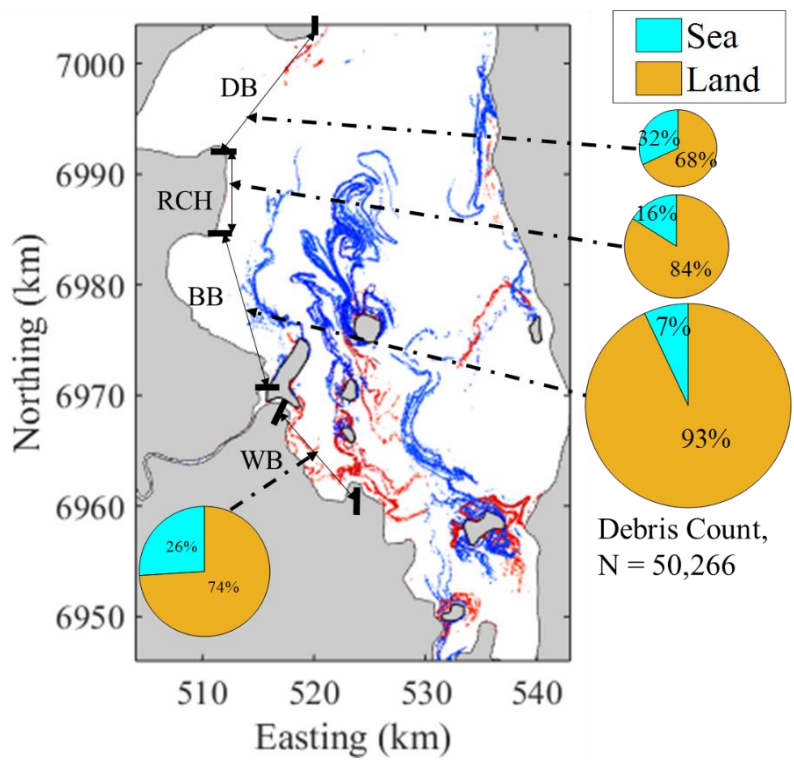

Figure 8. LCS in Moreton Bay superimposed with the distribution of marine debris collected along the shoreline: (DB: Deception Bay, RCH: Redcliffe Headland, BB: Bramble Bay, WB: Waterloo Bay): The area of the pie chart is represented by the scaled total number of debris collected along the indicated span of the shoreline during clean-up events from 2011–2018. $N = 50,266$ is the total number of debris collected from Bramble Bay indicated for scale reference.

The large percentage of sea-sourced debris collected in WB (26%) can be linked with the combination of the access to the Northern Passage provided between the LCS radiating from the Fisherman Island and Mud Island, combined with debris washed-up into MB through the Southern Passage. Another factor that can contribute to the increase in land-sourced debris is the differences in the population size and recreational water usage in these areas. This correlation is not examined in this study. The results here do however suggests that there is a link between source of debris and the visualised LCSs in complex tidal embayment. Extraction and prediction of LCS has a potential in the role of predicting debris sources and can be a useful way for effective source management plans for marine protected areas (MPA) such as MB.

### 3.4.2 LCS and fates of shoreline debris

The datasets for the debris clean-up within MB have shown the debris are predominantly land-sourced and exchange with the ocean is influenced by the LCS. Here, the fate of debris initially concentrated along the western shoreline in DB, BB and WB is examined. In order to quantify the extent of beaching or escape into the ocean for debris that were initially concentrated nearshore, particles were advected with the flow and wind fields using Eqs. 1, 4 and 5. In the analysis, the particles that are advected out of the domain through the land boundary are considered beached while particles escape to the ocean through the ocean boundaries. Particles are not reintroduced into the domain once they leave by either of the mechanisms because of the absence of datasets and appropriate parameterisation on this issue in the literature. Resuspension of beached particles as well as particle re-entrance during flood tides are important factors of consideration in a realistic debris fate models [*Critchell & Lambrechts*, 2016]. However, for the idealised case considered here, the fraction of particle that escaped into the ocean after 12 days was generally low (<20%). Therefore, the fraction that would re-enter into the domain later would be insignificant and is not included in this work.

The idealised case of debris fully submerged and perfectly locked to the surface flow. i.e. $A/W = 0$, where wind drag is negligible is first discussed. Figure 9 shows the initial distribution of the virtual tracers at DB, BB and WB. Their final locations just before exiting the flow domain or at the end of the advection time of 282 hours are shown in subsequent plots. To obtain tidally averaged statistics, the same initial distribution of virtual tracers was seeded to the flow at 24 different tidal phases from around low tide to the next low tide. The particle beached, averaged over a tidal cycle, as a function of time from release is presented in Figure 9(a). The results are presented for particles uniformly distributed with a spacing of 100 m, obtained from sensitivity analysis. Further increase in the particle density did not affect the statistics presented here. The results showed that <14% of tracer from WB were beached after 24 hours while 35% and 50% of tracers from BB and DB, respectively, were beached after 24 hours. At the end of 282 hours advection, 51% of particles initially deployed from WB were beached, 18% escaped into the ocean through the northern ocean boundary, while approximately 30% remained within MB, trapped along the island shores and stretched along the stable material line that connects with the shoreline adjacent north east of MB. In contrast to this pattern, > 95% of the particles deployed in the vicinity of DB and BB hit the shoreline before the end of the advection period and < 5% escaped into the ocean. The large beaching rate for these locations is consistent with the large land-sourced dataset collected from the clean-up events (Figure 8) and field surveys from other coastal estuaries [*Willis et al.*, 2017]. The small number of ocean escapes is consistent with the long residence time of 54–62 days within these bays [*Dennison & Abal*, 1999; *Gibbes et al.*, 2014]. The results here further indicate that the hydrodynamics of the north and south of MB are different.

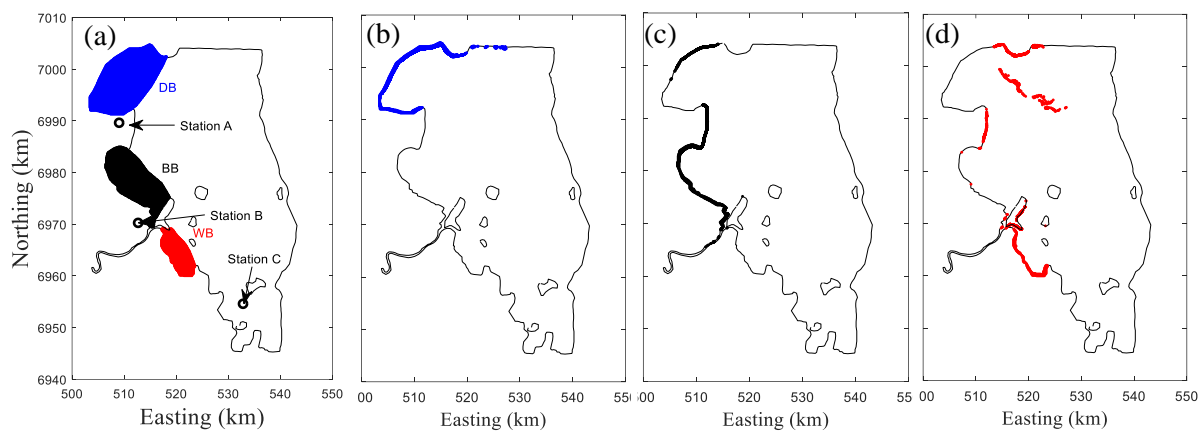

Figure 9: Fate of tracer particles initially uniformly distributed; (a) The initial distribution of tracer near the pockets of embayments with locations of weather stations (A, B and C; windroses are shown in Figure 10); (b-c) the locations of particles just before exiting the domain or final location after 282 hours for particles initially located in (b) Deception Bay, (c) Bramble Bay and (d) Waterloo Bay.

The result confirms that the initial location of tracer particles relative to the LCSs dictates, to a large extent, the fate of particles. In other word, particles trapped between the shoreline and the LCS have a higher chance of being beached while particles with clear pathway to the ocean through adjacent LCSs have lower chance of being washed ashore. For example, the tracer particles in the WB had clearer pathway in between the unstable material lines generated by the unsteady wake of Mud Island and Fisherman Island. This in combination with the obstruction provided by the islands explain why only a very small fraction (<1%) of the particles from WB ended up in the inner parts of BB and DB. The results highlight the role of LCSs in the fate of debris in the embayment and further confirm that the presence of the islands and headlands plays an important role in the distribution of debris and the physiochemical makeup of such an embayment.

We also consider the effect of both the initial tidal phase and the wind on fate of the nearshore tracer particles. Figure 10(a) shows the rate of tidal averaged percentage particles beached and Figure 10(b) shows the rate of escape of material into the ocean as a function of the initial tidal phase for the three bays (DB, BB, WB). The result showed that, consistent with tidal dynamics, the ocean escape is maximised for DB around high tide. However, maximum ocean escape rates were not particularly dictated by the tide for WB and BB.

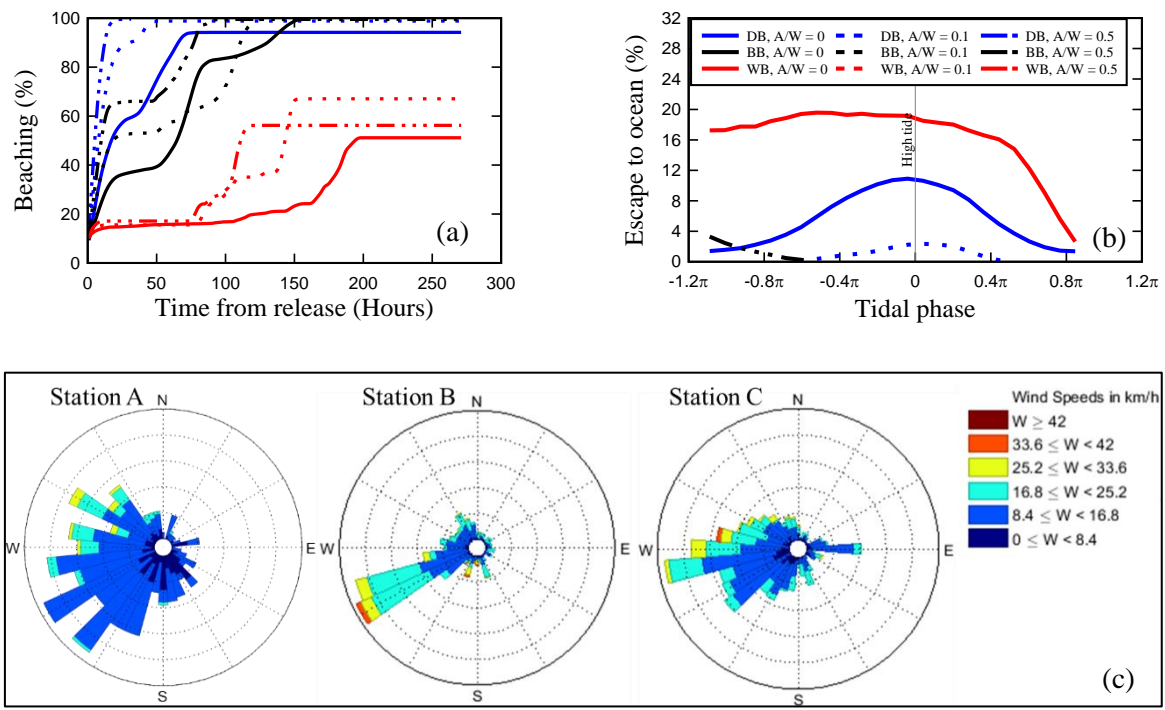

Figure 10: Fate of particles at selected embayments within Moreton Bay shown as (a) tidal averaged percentage tracer particles beached as a function of time; (b) percentage tracer particle escaped into the ocean after 282 hours as a function the initial tidal phase. (c) Wind roses for weather stations adjacent the bays (locations are shown in Figure 9a); Wind information covering the period particles were tracked; Wind adjusted to 10 m above sea level, direction showing where the wind is blowing from.

The role of wind on fate of debris is examined through the level of submergence $A/W = 0.0$, 0.1 and 0.5 covering submergence of different plastic material which dominates the debris washed up on Australia shoreline [*Hardesty & Wilcox*, 2011; *AMDI*, 2018]. In general, the wind was predominantly from the southwest (see wind roses in Figure 10c), favouring north-eastward material transport. The supplementary videos (SV3 and SV4) show the trajectories of the tracers with and without the wind drag included. The result is consistent with the net northward transport within MB and wind effect on the LCS. The rate of northward transport during the ebb tides increased while the rate of southwestward transport during the flood tide decreased. Overall, the wind effect caused a decrease in material beaching in the western shoreline and an increase in the material beaching in the northern shoreline. The wind drag reduced the rate of ocean escape close to <5%. This is consistent with wind effect on the unstable material lines which is shifted counter-clockwise towards the north and stretched out acting as a transport barrier for materials coming from the western shoreline.

## 4.0 Discussion

Hydrodynamic models currently provide fairly detailed and adequate information for the dynamics of tidal coastal embayments where radar data are unavailable. The dynamics of such systems are strongly time dependent on tide, wind, discharge from rivers/inlands and bathymetry [*Fischer et al.*, 1979; *Suara et al.*, 2019]. Here, the applicability of the concept of LCS using a FTLE diagnostic approach to describing the fate of floating materials in a coastal tidal embayment is demonstrated using MB as a case study.

There are some limitations in analysing Lagrangian structures from FTLE ridges. For example, *Haller* [2011] exposed some of the weaknesses associated with defining LCS by the ridges of FTLE field and showed that not all FTLE ridges are hyperbolic LCSs. In addition, two-dimensional analysis of three-dimensional flows does not necessarily reveal the 'true'

Lagrangian transport [*Branicki & Malek-Madani*, 2010; *Farazmand & Haller*, 2013]. Furthermore, hyperbolic structures computed from separate forward and backward integration of particles belong to different dynamical systems and do not evolve into each other in aperiodic flows [*Branicki & Malek-Madani*, 2010; *Farazmand & Haller*, 2013]. The transport barriers defined in the context of this work are not necessarily 'perfect'. They are rather regions in the domain with higher accumulation and spreading properties relative to others and are targeted at materials locked to the surface flow at a periodic tidal scale.

FTLE fields used as proxies to hyperbolic LCS highlighted the virtual structures responsible for the modification of transport of material at tidal scale. The relevant transport barriers are often attached to the islands and headlands along the shoreline. The results of the LCS analysis and parameters of the wake of the flow around the islands suggested that the interaction of the tidal flow with the islands generated unsteady wakes. These wakes leave a footprint of transport barriers in the FTLE field at an integration time of up to 72 hours. The results further showed that the northern part of MB with no islands exhibited different dynamics to the southern part which has multiple islands in close proximity to the shoreline. The relocation patterns of the LCSs were periodic with an oscillation period of ~12 hours reflecting strong tidal dynamics (Figure 6; Section 3.2). The structures were generally persistent with neap and spring tidal types. The comparison of the spatio-temporal distribution of the visualised LCS with the dataset from debris clean-up along the western shoreline in MB signalled that the relative locations of the LCS to the shoreline influence the beaching of debris (Figure 8).

The dynamical pictures of MB (Figures 5, 6 and 7) suggest a quantitative measure of mixing in the system. The presence of a repelling LCS indicates the proximity of a corresponding nearby attracting LCS. This is because an increase in eddy activity implies a simultaneous increase in both dispersion and attraction of material. Therefore, the information obtainable from the spatial average of the joint probability distribution of the forward and backward FTLE fields can be obtained from just one of the fields. Following *d'Ovidio et al.* [2004], the temporal variation of spatial averaged FTLE$^+$ which quantifies the mixing strength in the system is evaluated.

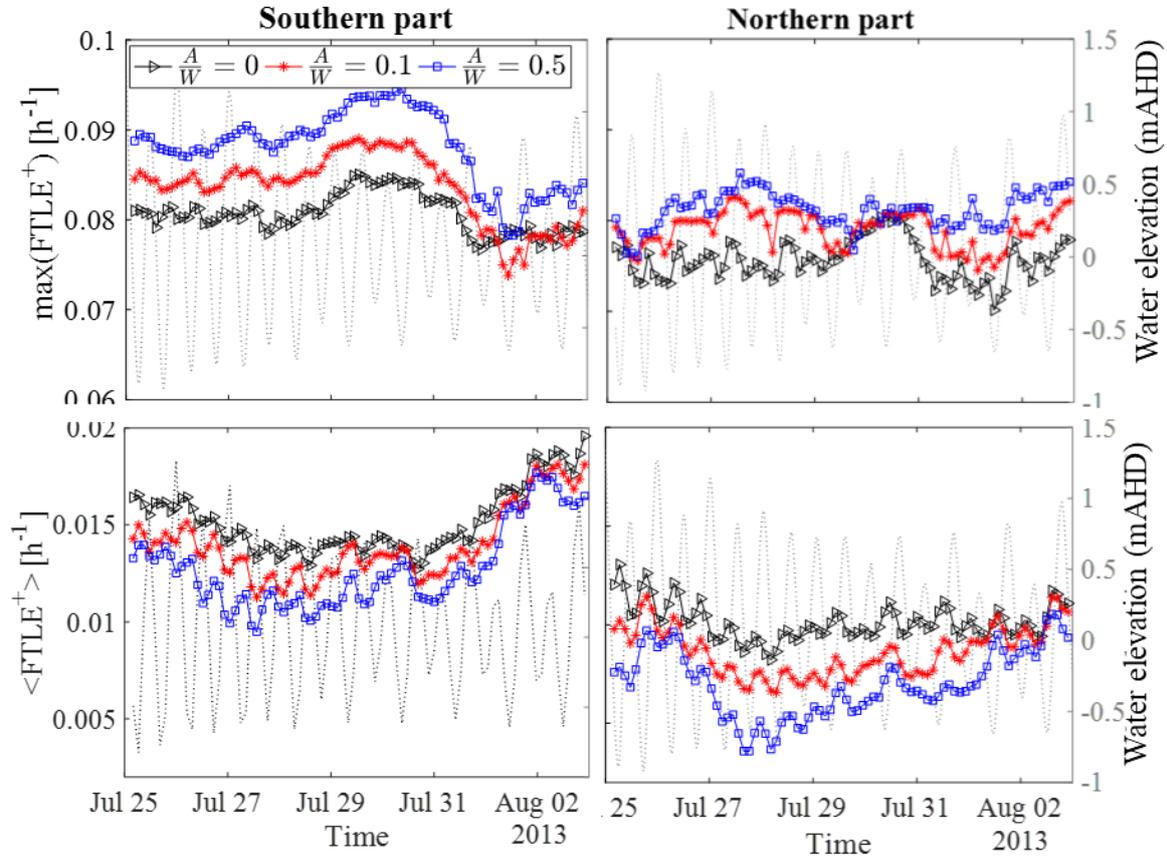

Figure 11: Effect of wind and tidal types on the overall mixing dynamics of the Northern and Southern parts of Moreton Bay. Upper panel: Maximum FTLE$^+$ field as a function of time. Lower panel: Spatial averaged FTLE$^+$ field as a function of time.

The calculations of the LCSs for the northern and southern parts were performed separately to examine the difference in their dynamics. This is combined with the temporal variation of the maximum FTLE field to provide the summary of the dynamics of the embayment for the 10 days of available velocity field. Although the spatial structures of the LCS are persistent and not significantly influenced by the tidal type, both the mixing strength and the rate of attraction toward the transport barriers varied as a function of time (Figure 11). The maximum value of the order of 0.075 h$^{-1}$ was attained, without a significant difference between the spring and neap tidal types, in the northern part. Maximum values of ~0.085 h$^{-1}$ were attained during the neap tides (29–31 July) with lower values obtained during the spring tides in the southern part. Conversely, the spatially average FTLE field indicated higher mixing activity during the spring tides compared to its level during the neap tides for both the northern and southern parts of MB. This implied that the rate of expansion/compression around the transport barriers were higher during the neap tide while mixing within the embayment was stronger for the spring tides.

The result showed that with the increase in windage, the rate of attraction toward visualised LCS in MB increased while the overall mixing strength reduced. The results also showed that inclusion of wind drag in the LCS analysis is important for predicting the fate of pollutant in such a tidal embayment. The wind stretched out the unstable material lines that were connected to the islands, radiating them north-westward and connecting them to the northern shoreline. This created a virtual barrier for transport of floating material emerging from the western shore and therefore wind is an important consideration for clean-up and pollutant management. Other effects including modification of debris pathways and fates in tidal waters due to the inertial size of debris, 3D flow dynamics, resuspension and re-entrance of debris not included here are still under investigation.

## 5.0    Conclusion

The study demonstrates the improved understanding of transport of debris in a complex tidal embayment using the Lagrangian coherent structures concept. The motivation for developing such an idea is the conceptual robustness of Lagrangian transport to motion of materials compared to that of a Eulerian concept and simplified trajectories of particles which are sensitive to model error, times and locations. The Lagrangian concept applied here particularly helps in understanding the fate of pollutants and the mixing strength in such systems. It also unveiled the likely pathways for floating material transport, important information for guiding at-sea clean-up of debris and management of accidental floating material spillage. We performed the analysis in which potential LCSs are visualised from ridges of the FTLE field obtained from forward and backward integration of virtual particles. The results showed that tidal scale motion, pathways and fate of floating materials, are affected by the interaction of the tidal flow and physical structures in such an embayment. Islands and headlands in the embayment were shown to play an important role in the transport and mixing of debris.

The analyses highlight the significant role of the wind field in modifying the location of the visualised LCS structures as well as the hyperbolic saddle points responsible for mixing in such a system. The period was selected based on the available validated hydrodynamic model output. Importantly, the analysis here showed that the orientation of the visualised LCSs explains the dominant land-sourced debris from clean-up datasets and the high residence time (>50 days) in the western part of the embayment as reported in the literature. The results signal that there is a link between the source of debris and LCSs, therefore, extraction of LCSs can help to predict pollutant sources and thus, be a useful way for effective Source Reduction Plans for marine protected areas such as MB.

## Acknowledgements

The authors thank the Australian Marine Debris Initiative, the community organisations and individuals involved in the collection and provision of the data used in this report. The project is supported through Australian Research Council Linkage Project grant LP150101172 and Discovery Project grant DP190103379. TS acknowledges support from Estonian Research Council grant IUT33-3 and from the Estonian Research Infrastructures Roadmap object Info-technological Mobility Observatory (IMO). The authors acknowledge the contribution of Professor Hubert Chanson for advice and discussion on the work.